\documentclass{article}
\usepackage{spconf, graphicx, amsmath, hyperref,setspace,enumitem}
\usepackage{subcaption}
\usepackage{multirow}

\newcommand\figref{Figure~\ref}
\newcommand\tabref{Table~\ref}

\input{def.set}

\title{Learn to Remix Rather Than Separate:\\ End-to-end neural remixing with joint optimization}

\title{Don't Separate, Learn to Remix:\\ End-to-End Neural Remixing with Joint Optimization}
\name{Haici Yang$^1$, Shivani Firodiya$^1$, Nicholas J. Bryan$^2$, Minje Kim$^1$}
\address{ $^1$Indiana University, Luddy School of Informatics, Computing, and Engineering, Bloomington, IN, USA\\
  $^2$Adobe Research, San Francisco, CA, USA}

\ninept
\begin{document}

\maketitle

\begin{abstract} 
The task of manipulating the level and/or effects of individual instruments to recompose a mixture of recordings, or remixing, is common across a variety of applications such as music production, audio-visual post-production, podcasts, and more. This process, however, traditionally requires access to individual source recordings, restricting the creative process. To work around this, source separation algorithms can separate a mixture into its respective components. Then, a user can adjust their levels and mix them back together. This two-step approach, however, still suffers from audible artifacts and motivates further work. 
In this work, we learn to remix music directly by re-purposing Conv-TasNet, a well-known source separation model, into two neural remixing architectures. To do this, we use an explicit loss term that directly measures remix quality and jointly optimize it with a separation loss.
We evaluate our methods using the Slakh and MUSDB18 datasets and report remixing performance as well as the impact on source separation as a byproduct. Our results suggest that learning-to-remix significantly outperforms a strong separation baseline and is particularly useful for small volume changes.
\end{abstract}


%



\section{Introduction}\label{sec:introduction}

Remixing or the task of manipulating the level and/or effects of individual instruments to create a derivative recording is widely used for audio and music production applications such as music content creation (e.g. DJ performances), audio-visual post-production, remastering, podcasting, and more. 
Music remixing, in particular, is of critical interest and can be used to modify an original version of a song into a different version 
to suit a specific genre, e.g., from country to rock; or to alter the sound stage, e.g., re-position an instrument's stereophonic location from the center to the left.  

In this paper, we focus on the application where a user wants to \textit{boost} or \textit{suppress} arbitrary instruments differently. This is a significant challenge for a computer algorithm 
as those multiple sound sources are recorded altogether as a single mixture. 
Approaches to solve this problem can be categorized into two kinds: feature-based methods \cite{YoshiiK2005inter, GilletO2005extraction} and music source separation-based methods \cite{WoodruffJ2006remixing, ItoyamaK2008instrument}. Feature-based remixing systems report successful performance in both attenuation ($-10$ and $-6$ dB) and amplification ($10$ dB) tasks, however, they only work on a small number of specific instruments, thus limiting usability. 
In contrast, music source separation (MSS) algorithms allow users to manipulate estimates of multiple separated instruments to achieve remixing. 
More specifically, source separation and remixing are treated as two independent processes, where remixing serves as a post-processor that completely relies on the separation effect.
This makes remix quality highly dependent on MSS performance. Recent progress in MSS mostly happens with deep learning methods \cite{Manilow2020hierarchical,Brocal2020content,StoterF2019open} on both frequency-domain \cite{HuangP2015ieeeacmaslp,UhlichS2017improving,SeetharamanP2019class,LeeJ2019audio,SeetharamanP2019bootstrapping,HersheyJ2016icassp} and time-domain \cite{StollerD2018waveunet,Lluis2019,DefossezA2021mss,LuoY2019conv-tasnet}. Meanwhile, manipulating and interacting in the latent space have proven to be beneficial for source separation and related problems \cite{YangH2021sanac,BryanN2013icml}.



\begin{figure}
\centering
        \includegraphics[width=.57\columnwidth]{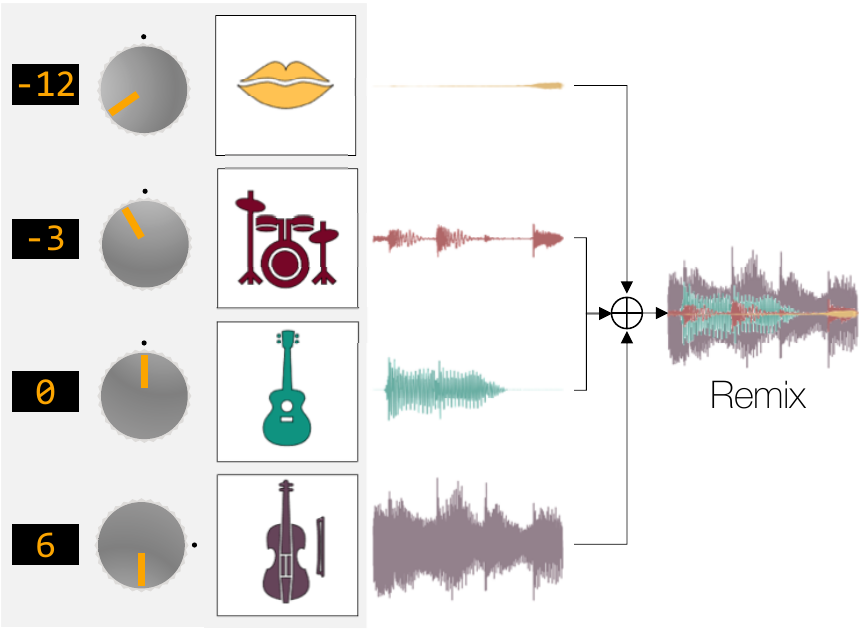}
        \caption{We perform end-to-end remixing by jointly optimizing for music source separation and remixing loss. 
        }
        \label{fig:gui}
\end{figure}


In our proposal, we aim to minimize separation artifacts found in remixing systems by learning to remix directly. To do so, we first justify our problem formulation with analysis on commonly used source separation evaluation metrics. Then, we extend one of the state-of-art source separation models, Conv-TasNet \cite{LuoY2019conv-tasnet}, and propose two adaptations toward end-to-end remixing (a) one that applies remixing weights to the source estimates (b) a latent variable control by applying the weights to the bottleneck feature of Conv-TasNet, which is why we chose it as the baseline model. Both of them are regularized, so the system is optimized to perform separation and remixing simultaneously. We evaluate the proposed methods on two music source separation datasets, Slakh \cite{ManilowE2019slakh} and MUSDB18 \cite{musdb18-hq}, and analyze the behavior of the models on various remixing scenarios.
To summarize, we make the following key contributions:
\begin{itemize}[noitemsep,topsep=0pt, leftmargin=0in, itemindent=.15in]
    \item To the best of our knowledge, we propose the first end-to-end neural method that jointly learns MSS and remixing together. 
    \item We show  our proposed neural remixing method is capable of a wider range of volume change compared to existing methods, ranging from $-12$ to $12$ dB, and can deal with up to five sources.  
\end{itemize}

\begin{figure}[t]
        \centering
        \includegraphics[width=0.95\columnwidth]{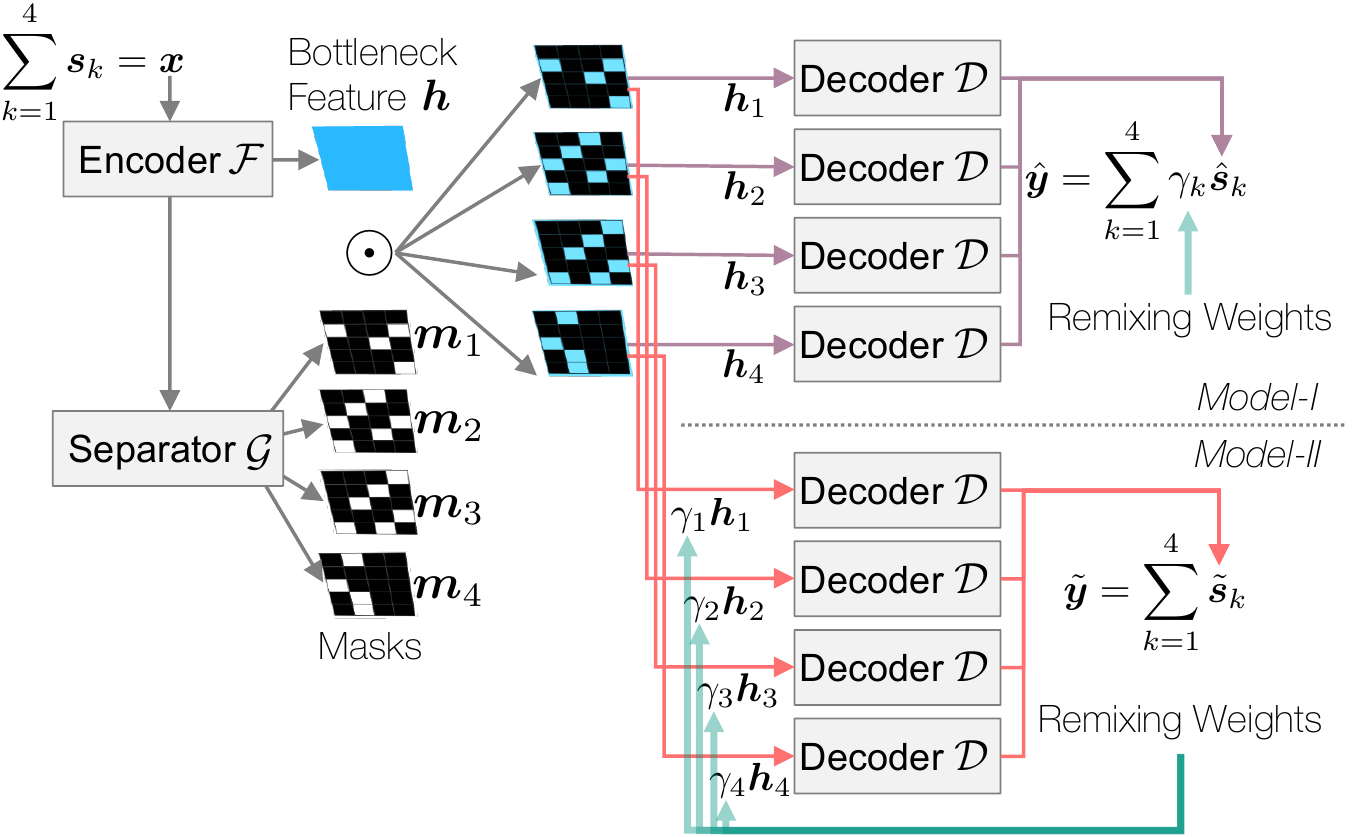}
        \caption{Our neural remixer architectures.  \textit{Model-I} jointly optimizes a separation and remix loss. \textit{Model-II} extends this further by performing remixing in the model latent space.}
        \label{fig:model}
\label{fig:architecture}
\end{figure}

\section{Problem Formulation}\label{sec:problem}
We argue that the na\"ive concatenation of the separation and remixing processes is artifact-prone. In a two-source case, for example, we decompose the recovered source into three components:
\begin{equation}\label{eq:decomp}
    \hat\bs_1 = \alpha_1\bs_1+\beta_1\bs_2+\bolde_1,\quad\hat\bs_2 = \alpha_2\bs_2+\beta_2\bs_1+\bolde_2.
\end{equation}
The reconstruction of the first source $\hat\bs_1$, for example, consists of the scaled ground-truth $\alpha_1\bs_1$, scaled interference $\beta_1\bs_2$, and the artifact generated during the separation process $\bolde_1$. Hence, the perfect scale-invariant separation is achieved when $\beta_1, \beta_2=0$ and $\bolde_1, \bolde_2=\bzero$. This decomposition is common in source separation evaluation \cite{VincentE2006ieeeaslp}. 

Here, we introduce a pair of non-negative remixing weights $\gamma_1$ and $\gamma_2$ as the intended scaling factors. The estimated remix $\hat y$ is derived by the reconstruction of sources, which can be rearranged by using Eq. \eqref{eq:decomp}:
\begin{equation}\label{eq:remix_artfacts}
    \begin{split}
        \hat y &= \gamma_1\hat\bs_1+\gamma_2\hat\bs_2\\&=(\gamma_1\alpha_1+\gamma_2\beta_2)\bs_1
        +(\gamma_1\beta_1+\gamma_2\alpha_2)\bs_2+\gamma_1\bolde_1+\gamma_2\bolde_2.
    \end{split}
\end{equation}

Clearly, imperfect separation can cause compromised weighting because the approximation $\gamma_1\approx \gamma_1\alpha_1+\gamma_2\beta_2$ is inaccurate if $\beta_2$ is too large or $\alpha_1$ is too small. Furthermore, the artifact $\gamma_1\bolde_1+\gamma_2\bolde_2$ is not guaranteed to cancel each other after scaling.  Thus, we seek an alternative approach.

\section{Methodology}\label{sec:method}

We propose to jointly learn source separation and remixing objectives. Although combining multiple sources' and mixture's reconstruction terms have been used to improve source separation performance \cite{ChoiHS2018phase-aware,SawataR2020cumx}, they are not in the context of remixing as in our proposed method.
Let $L_k$ be the user's intended volume change amount for source $k$ in terms of sound pressure level (SPL) relative to the original loudness,
$L_k=10\log_{10} (\gamma_k\bs_k)^\top(\gamma_k\bs_k) / \bs_k^\top\bs_k$,
where $\gamma_k$ denotes the corresponding remixing ratios for amplitude. For example, $L_k=+10$ dB is converted into $\gamma_k=3.16$.
For a $K$-source mixture, the remix target is defined as $\by=\sum_{k=1}^K \gamma_k\bs_k$, while the input mixture is $\bx=\sum_{k=1}^K\bs_k$. 


\subsection{The baseline: remixing estimated sources}

Our baseline utilizes Conv-TasNet as a source separation module
that takes time-domain signals as input and computes the loss in the time domain as well, i.e., $\calL_\text{BL} = \sum_{k=1}^K \calE(\bs_k||\hat\bs_k)$, 
where we use signal-to-noise ratio (SNR) for the error function $\calE(\cdot||\cdot)$. Then, the user's intended source-specific scales $\gamma_k$ are applied to the source estimates to approximate the remix with $\hat\by$:
    $\by\approx\hat\by = \sum_{k=1}^K \gamma_k\hat\bs_k$. 
Note that this process is prone to inaccuracy in source control and artifacts as shown in eq. \eqref{eq:remix_artfacts}. 

\subsection{\textit{Model-I}: the proposed remixing loss}

We propose incorporating the remixing process in training---synthesizing the remix using the recovered sources, and comparing it to the target remix to compute the loss. Hence, our \textit{Model-I} incorporates a loss that penalizes insufficient remix quality as well as the source-specific reconstruction:
\begin{equation}\label{eq:model1_loss}
    \calL_\text{\textit{Model-I}} = \psi\calE(\by||\hat\by)+\lambda\sum_{k=1}^K \calE(\bs_k||\hat\bs_k), 
\end{equation}
where $\psi$  and $\lambda$ control the contribution of the remix and source-specific reconstruction losses, respectively. The rationale behind our proposed loss is to calibrate the optimization results of source separation by imposing a cost on the remix quality. 
We also believe this change can help the system balance the SIR-SAR trade-off for a better remix quality.



\subsection{\textit{Model-II}: the proposed control of the latent space}

Our second proposed approach is to control the loudness of sources in the latent space. 
Unlike \textit{Model-I}, \textit{Model-II} applies the remixing weights $\gamma_k$ to one of the network's hidden layers, instead of source estimates as shown in Fig. \ref{fig:model}.
Conv-TasNet provides a convenient framework to do this. In Conv-TasNet, the separation is conducted first by computing the bottleneck feature map via the \textit{encoder} module $\calF(\cdot)$ and  $K$ masks using the \textit{separator} module $\calG(\cdot)$, 
\begin{equation}
    [\boldm_1, \boldm_2, \ldots, \boldm_K]\leftarrow\calG(\bx),\quad\bh\leftarrow\calF(\bx),
\end{equation} 
where the masks are probabilistic, i.e., $\sum_{k=1}^K \boldm_k=\bone$. Then, the bottleneck feature $\bh$ is distributed to $K$ different source-specific feature spaces via masking, which are then decoded back into the time-domain  using the \textit{decoder} module $\calD(\cdot)$, respectively:
\begin{equation}
    \bh_k\leftarrow \boldm_k \odot \bh, \quad \hat\bs_k\leftarrow\calD(\bh_k), \quad\forall k,
\end{equation} 
where $\odot$ denotes Hadamard product. 


By making use of the separated hidden space, \textit{Model-II} modulates the hidden variables by multiplying their corresponding remixing weights $\gamma_k$:
    $\tilde{\bs}_k\leftarrow\calD(\gamma_k\bh_k)$,
where the output $\tilde{\bs}_k$ attempts to reconstruct scaled sources $\gamma_k\bs_k$ directly. \textit{Model-II}'s loss is then
\begin{equation}
    \calL_\text{\textit{Model-II}}=\psi\calE(\by||\tilde{\by})+\lambda\sum_{k=1}^K \calE(\gamma_k\bs_k||\tilde{\bs}_k) , 
\end{equation}
where $\tilde{\by}=\sum_{k=1}^K \tilde{\bs}_k$. 
What distinguishes \textit{Model-II} from \textit{Model-I} is that  \textit{Model-II} has the ability to associate the separation behavior with the remix weights during inference.  We believe this can help the decoder handle any additional artifacts or interference that is introduced. 

For example, if a user wants to boost $j$-th source while suppressing others, $\gamma_j \gg \gamma_k,~\forall k\neq j$, our models can focus more on the precise reconstruction of the dominant source $\bs_j$ than the other sources that will be suppressed anyway. Similarly, $\gamma_k\bolde_k$ is small to ignore when $k\neq j$, while suppressing $\bolde_j$ is important as $\gamma_j$ is large. Another important corner case is when the remix target is very similar to the input, i.e., $\by\approx\bx$ when $\gamma_k\approx1,~\forall k$. In this trivial case, the proposed models can save unnecessary separation effort.

\begin{table*}[t]
  \centering
  \resizebox{.9\textwidth}{!}{%
  \begin{tabular}{c|c|c|c|c|c|c|c|c}
  \noalign{\hrule height 1pt}
    \multicolumn{2}{c|}{\textit{minSDR} / LD} & Baseline & \multicolumn{3}{c|}{\textit{Model-I}} & \multicolumn{3}{c}{\textit{Model-II}}  \\
    \hline
    Train + Test & $K$ & $\psi:\lambda=0:1$ & $\psi:\lambda=1:1$ & $\psi:\lambda=K:1$ & $\psi:\lambda=1:0$ & $\psi:\lambda=1:1$ & $\psi:\lambda=K:1$ & $\psi:\lambda=1:0$ \\
    \noalign{\hrule height 1pt}
    \multirow{4}{*}{\shortstack{Slakh \\+\\ Slakh}} & $2$ & 28.24 / 0.18 & 24.59 / 0.31 & 27.63 / 0.21 & \textbf{28.84 / 0.19} & 27.35 / 0.19 & 28.34 / 0.21 & 27.16 / 0.19 \\
    & $3$ & 18.72 / 0.67 &19.88 / 0.8 &19.7 / 0.87& \textbf{21.26 / 0.67}& 20.09 / 0.69&19.81 / 0.77 & 19.26 / 0.81 \\
    & $4$ & 0.22 / 8.42 & 16.48 / 1.54 &15.24 / 1.85 & 15.57 / 1.72 & 16.8 / 1.57 & 15.16 / 1.79 & \textbf{17.23 / 1.51}\\
    & $5$ & -4.08 / 11.31 & 7.92 / 3.87 & 12.2 / 3.2 & 11.71 / 3.34 & 8.24 / 3.86 & \textbf{12.44 / 3.15} & 11.5 / 3.45 \\
    \hline    
    \multirow{3}{*}{\shortstack{MUSDB18 \\+\\ Slakh}} & $2$ & 23.83 / 0.35 & 23.19 / 0.47 & 23.01 / 0.45 & 24.96 / 0.39 & 23.99 / 0.44 & 23.97 / 0.41 & \textbf{25.15 / 0.35} \\
    & $3$ & 11.88 / 1.64 & 14.13 / 1.72 & 13.37 / 1.94 & \textbf{15.3 / 1.6} & 15.2 / 1.56 &14.76 / 1.49 & 15.15 / 1.68\\
    & $4$ & -6.06 / 7.85 & 9.74 / 2.78 & 9.94 / 2.8 & 9.19 / 3.05 & 9.63 / 2.88 & \textbf{10.2 / 2.78} & 9.73 / 3.01 \\
    \hline    
    \multirow{3}{*}{\shortstack{MUSDB18 \\+\\ MUSDB18}} & $2$ & 17.33 / 0.92 & 17.55 / 0.98 & 17.28 / 0.88 & 18.08 / 0.95 & 17.7 / 0.96 & 17.87 / 0.84 & \textbf{18.13 / 0.97} \\
    & $3$ & 11.82 / 1.94 & 13.37 / 1.93 & 12.52 / 2.29 & \textbf{14.49 / 1.72} & 14.17 / 1.71 & 14.13 / 1.64 & 14.15 / 1.94 \\
    & $4$ & -9.16 / 10.1 & 10.16 / 2.93& \textbf{11.01 / 2.85} & 9.84 / 3.26 & 10.49 / 2.97& 10.95 / 3.0 & 10.01 / 3.27 \\
    \hline
    \multirow{3}{*}{\shortstack{Slakh \\+\\ MUSDB18}} & $2$ & 12.26 / 1.61 & 14.54 / 1.31 & 14.54 / 1.39 & 14.71 / 1.36 & 14.25 / 1.42 & \textbf{15.1 / 1.29}  & 13.43 / 1.56 \\
    & $3$ & 8.27 / 2.59 & 9.37 / 2.85 & 10.16 / 2.73 & 10.21 / 2.75 & 9.69 / 2.72 & 10.18 / 2.62 & \textbf{10.57 / 2.48} \\
    & $4$ & -6.33 / 9.88 & 7.46 / 3.77 & \textbf{8.44 / 3.66} &  8.34 / 3.65 &  7.75 / 3.68 & 8.29 / 3.68 &  8.06 / 3.76 \\
    \noalign{\hrule height 1pt}
\end{tabular}}
  \caption{Cross-dataset evaluation on \textit{minSDR} and loudness difference (LD) for re-mixture construction.}
  \label{tab:bigone}
\end{table*}

\section{Experiments}\label{sec:Experiment}

\subsection{Datasets}

We use two music datasets: Slakh \cite{ManilowE2019slakh} and MUSDB18 \cite{musdb18-hq}. While Slakh consists of a large number of stem tracks synthesized using virtual instruments, the MUSDB18 samples are real professional-level recordings. However, the quantity of MUSDB18 is often not enough to train a large model that generalizes well to  real-world music signals.
We first train and test within each dataset independently. Cross-database testing follows, i.e., train on Slakh and test on MUSDB18, and vice versa.
In MUSDB18, the sources consist of \texttt{vocals}, \texttt{drums}, \texttt{bass}, and \texttt{other}. For two-source experiments, for example, the mixture consists of \texttt{vocals} and \texttt{drums}. As for Slakh, we test up to five sources in the following order: \texttt{piano},  \texttt{drums}, \texttt{bass}, \texttt{guitars} and \texttt{string}.


\subsection{Training setup}
We build our models and experiments off of Asteroid's Conv-TasNet implementation \cite{ParienteM2020asteroid}.  We fix the order of the sources and do not use permutation invariant training (PIT) \cite{YuD2017pit}. We use the entire training set for training ($\sim$48h for Slakh and $\sim$6h for MUSDB18), and split the original test set ($\sim$12h and 3h respectively) into validation and test sets evenly.  The model is trained on one-second segments  until the validation loss does not improve in 50 consecutive epochs. Only segments with all instruments active are included during training as we found this was better than having segments with silent instruments in initial experiments. For evaluation, no such restriction is applied. Adam is used as the optimizer with an initial learning rate of $1\times 10^{-3}$ \cite{KingmaD2015adam}. During training, the $K$ remix weights are sampled randomly from the range between $-12$ to $12$ dB and used to form the target remix segments. All signals are sampled at $44,100$ kHz.


Note, we substitute the original scale-invariant signal-to-distortion ratio (SI-SDR) loss function \cite{LeRouxJL2018sisdr} in Conv-TasNet with classical source-to-noise ratio (SNR) (equals to \texttt{bss\_eval\_images}’s SDR \cite{VincentE2006ieeeaslp}) to suit the scale-sensitive problem. The scale-dependent SDR (SD-SDR) proposed together with SI-SDR is potentially another choice for general scaling control. However, the nature of SD-SDR is that it is insensitive to large up-scaling factors, while our model is designed to tackle both downside and upside scales. We also did a pre-experiment on training with SDR or SD-SDR as the loss function, and results support our choice for SDR. 

\subsection{Evaluation setup}
For evaluation, we explicitly evaluate the remix task. To do this, we synthesize ground-truth remix mixtures and then compare our estimated remix mixtures to this ground truth. To create the remixes, we manipulate one source at a time ranging from $-24$ to $+24$ dB with a step size of $3$ dB, totaling 17 discretized values, and sum the result. This  range is twice wider than used for training.  We repeat the experiment for all $K$ sources, individually. 
Note, while we focus on the common one-source manipulation case, our models can control multiple sources at the same time. 


We objectively assess our models based on the two criteria suggested in \cite{WierstorfH2017perceptual}, sound quality and loudness balance.
Given that both the up-scaling and down-scaling in our problem is critical, we use the minimum of SDR and SD-SDR using the estimated and ground-truth remix mixtures as a measure for sound quality \cite{LeRouxJL2018sisdr}, which we denote as \textit{minSDR}.  
To evaluate the loudness balance, we consider the sources re-mixed in a linear time-invariant manner, and decompose the estimated re-mixture using a least square algorithm. The coefficients obtained for each source in the process can be viewed as the actual scaling factors that have built the estimated re-mixture. Given the output and target scalars, we finally report the loudness difference in decibels. Finally, we additionally compute the SIR and SAR scores using \texttt{bss\_eval\_images} to investigate the impact of remixing on the separation behavior.
\section{Results and Discussion}\label{sec:Results}

\subsection{Average remix performance}
We report the mean and standard deviation of SDR values of the estimated remix compared to the ground truth in \tabref{tab:bigone}. Overall, both our proposed models achieve a better remix quality than the baseline, and the improvement is more significant as the number of sources increases. The same trend can be found in the cross-database testing cases. There is no difference between \textit{Model-I} and \textit{II}. 

When we look at the merit of our remix loss controlled by $\psi$, we find: 1) the remix-only loss ($\lambda:\psi=0:1$) is preferred in all but one $K=2$ and $K=3$ cases 2) when the number of  sources increases or the task gets harder (e.g doing the cross-database test), it is beneficial to have higher $\lambda$ for better separation control, i.e. $\lambda:\psi=1:4$, and 3) the proposed loss causes the significant performance gap between the baseline and proposed models, as the baseline model uses $\psi=0$.

\begin{figure*}[t]
     \centering
     \begin{subfigure}[b]{0.34\textwidth}
          \includegraphics[width=\textwidth]{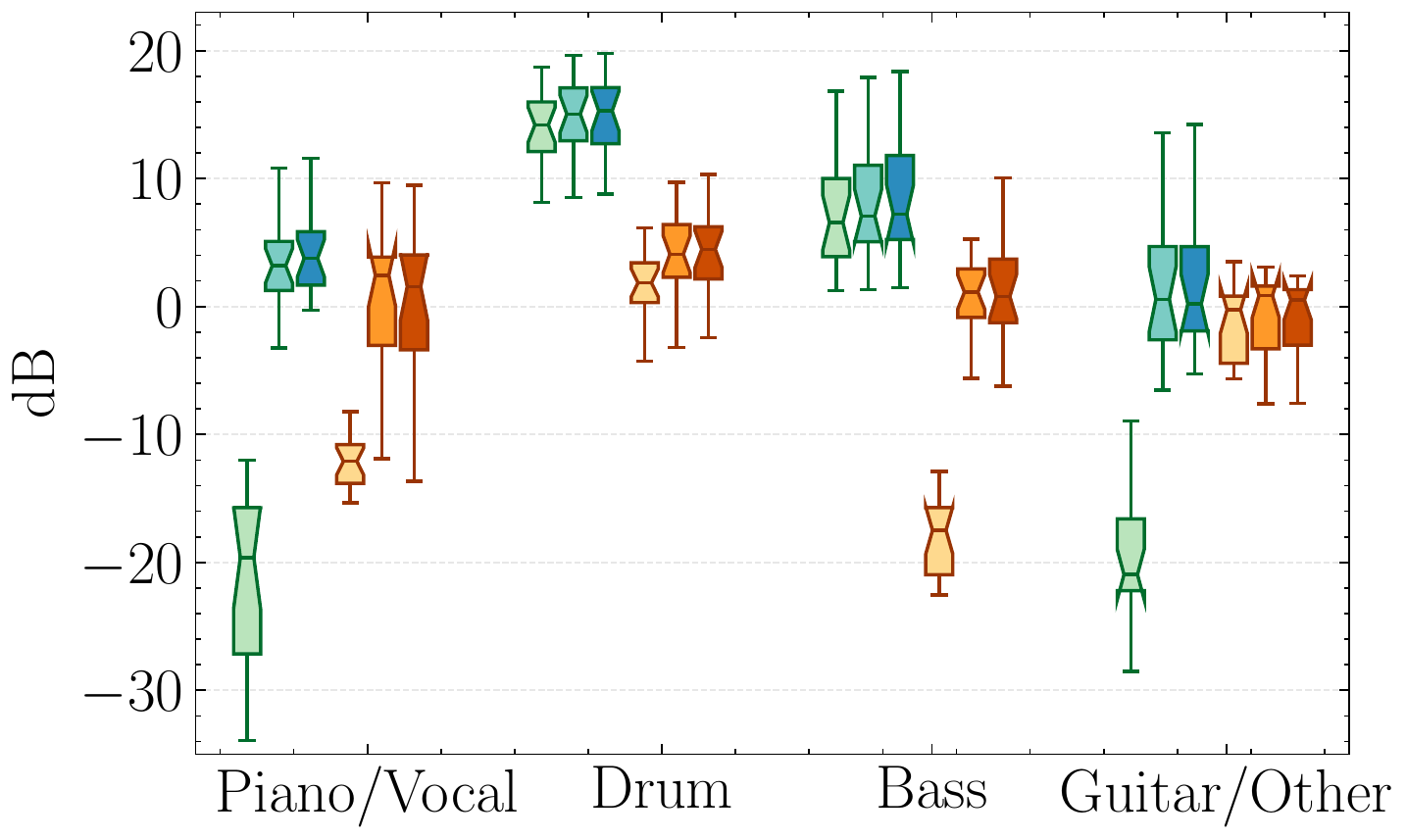}
         \caption{SDR}
     \end{subfigure}
     \begin{subfigure}[b]{0.325\textwidth}
          \includegraphics[width=\textwidth]{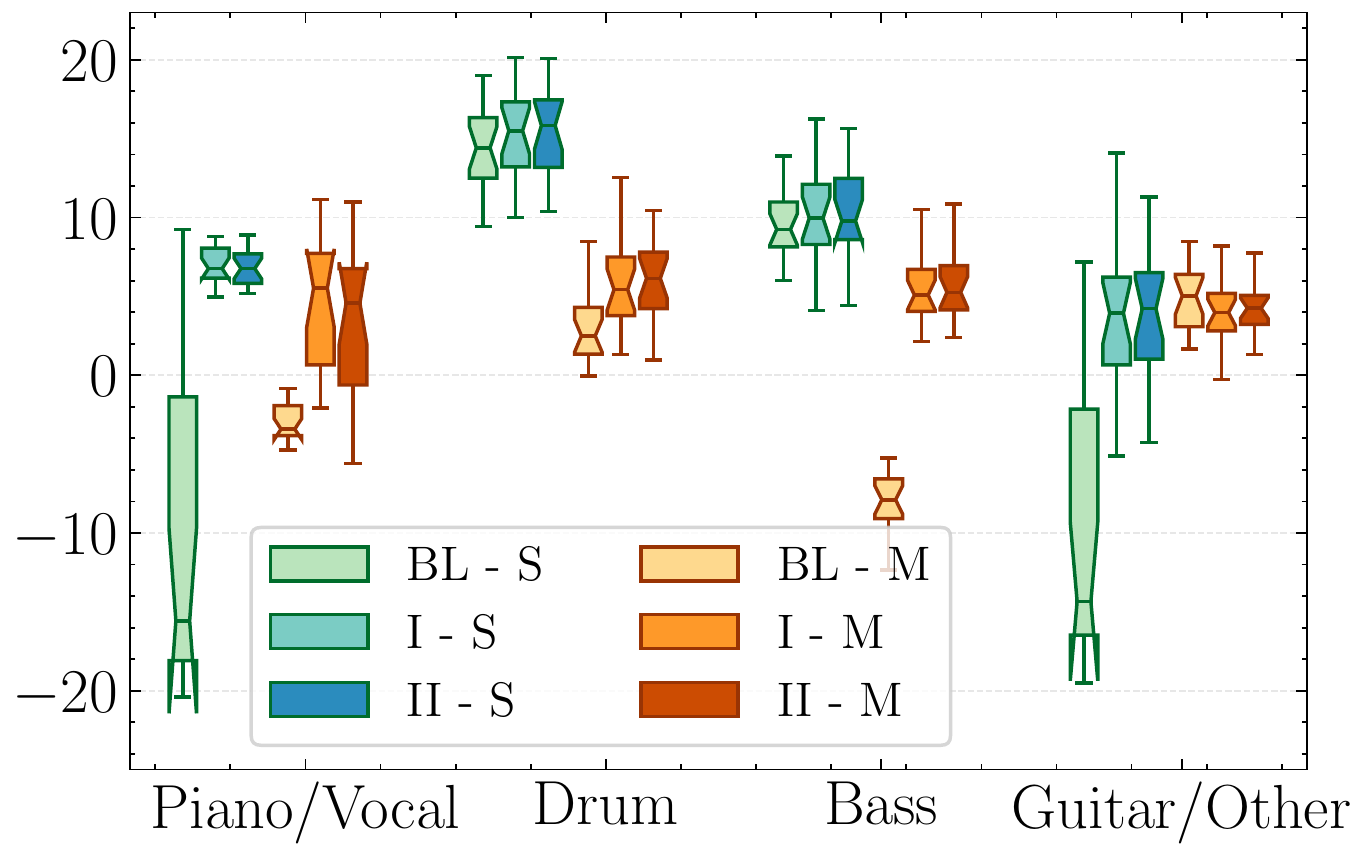}
         \caption{SAR}
     \end{subfigure}
     \begin{subfigure}[b]{0.325\textwidth}
          \includegraphics[width=\textwidth]{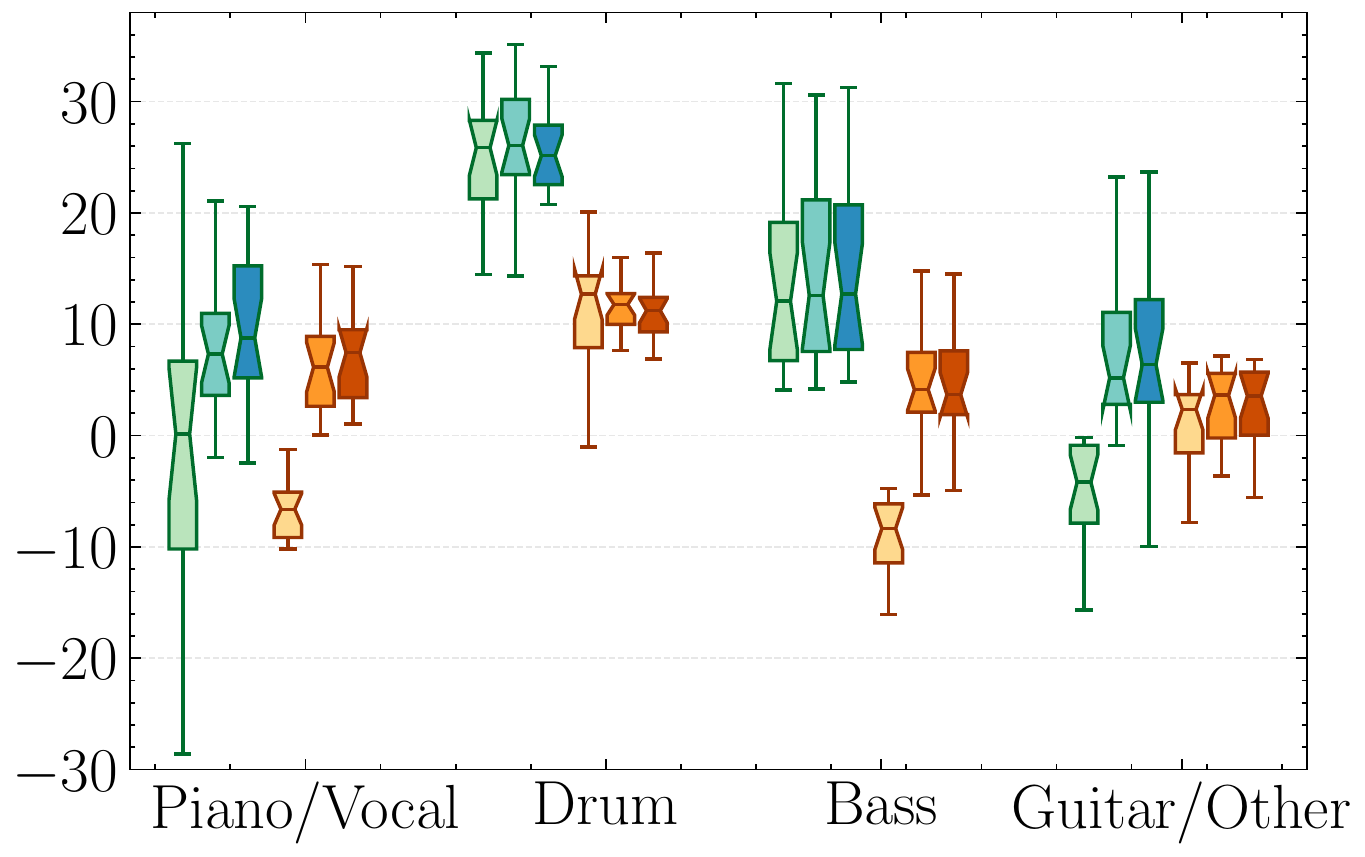}
         \caption{SIR}
     \end{subfigure}
    \caption{Source separation scores of the baseline (BS), \textit{Model-I} (I), and \textit{Model-II} (II) on Slakh (S) and MUSDB18 (M) for the four-source cases. The first and last box groups are \texttt{piano} and \texttt{guitars} in the Slakh experiments, and \texttt{vocals} and \texttt{others} in the MUSDB18 case. }
    \label{fig:bss}
     \vspace{-4mm}
\end{figure*}



\subsection{Vocal remix vs. loudness performance}


\begin{figure}[t]
     \centering
    \begin{subfigure}[b]{0.49\columnwidth}
          \includegraphics[height=.9in]{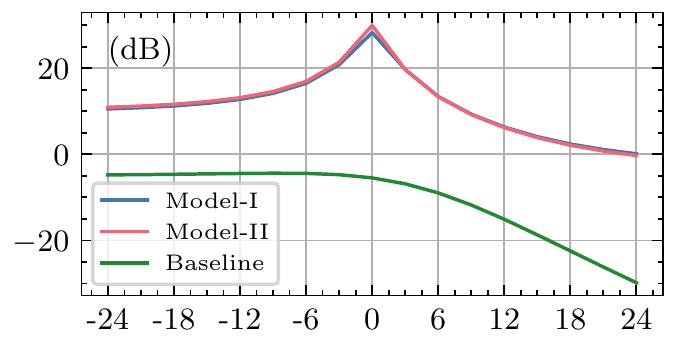}
    \end{subfigure}
    \hfill
    \begin{subfigure}[b]{0.49\columnwidth}
          \includegraphics[height=.9in]{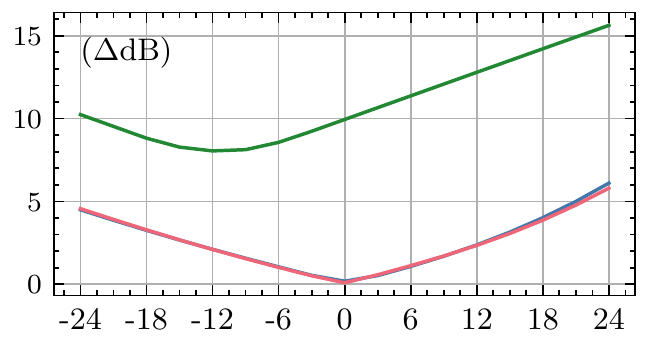}
     \end{subfigure}
    \caption{Remix \textit{minSDR} (left) and loudness difference (right) with respect to vocal volume adjustment.}
    \label{fig:ratio_control}
     \vspace{-4mm}
\end{figure}

In \figref{fig:ratio_control}, we display the change of remix quality and loudness difference along with the extent of vocal volume control in the four-source cases using MUSDB18. The x-axis represents adjusting the level of vocal from $-24$ to $+24$ dB, while the other instruments are kept unchanged. This intends to mimic the scenario where the user can adjust the volume of one instrument source by twisting a knob. 


\figref{fig:ratio_control} clearly shows that our proposed models outperform the baseline consistently in all the remixing weight choices, with higher reconstruction score and lower loudness difference. The trend is especially salient if the volume control amount is near $0$ dB, that is when the intended remix is similar to the input mix. It is noticeable that the baseline's 
\textit{minSDR} score decreases almost monotonically as the remix weights change, and the minimum loudness difference for the baseline model deviated from $0$ dB. These patterns indicate that, when the target remix stays around the original mix, our proposed models show more advantage as they can effectively focus on the artifacts rather than trying to suppress interference. Note that \textit{Model-I} addresses this issue only during training, while \textit{Model-II} is capable of reflecting the remixing weights to the decoding process during testing. Meanwhile, for the baseline model, the artifact and interference contained in the source estimates can be heterogeneous, making their remix stand out as distortion. Therefore, monotonic control of a source estimate can consequently monotonically influence the total remix quality, and the lowest loudness difference is reached when depressing the source volume and the distortion with it. 



Because of the reason stated above, our proposed models' performance changes are more predictable---the performance graphs share a similar pattern in all experiments on various instruments, datasets, and both proposed models, providing a stable user experience. 
Their asymmetric shape is caused by our setup where we boost or suppress only one source: boosting tends to exhibit more artifacts. 
In contrast, the baseline's performance is less predictable. For some sources, volume amplification has a negative impact while others suffer from a reduced volume. This observation echoes the baseline's behavior reported in \figref{fig:bss}, where performance varies a lot over the choice of source. 



It is interesting to note that, although \textit{Model-I} does not have any inference-time mechanism to adjust the separation behavior according to the different remixing weights, it still performs on par with \textit{Model-II}. We believe that to achieve a good sound quality for remixing, it is most crucial to reduce the artifacts produced in the separation step, and these two proposed models almost reach the same level in pursuing this goal. 
However, we believe that \textit{Model-II}'s source control in the latent space is a potentially more useful approach to more complex remixing tasks beyond volume adjustment such as source-specific nonlinear filtering.  

\subsection{Music source separation performance}
To investigate the impact of our remix loss on the separation behavior, we compute the SDR, SIR and SAR of the three models' separation results using the \texttt{BSS\_Eval} toolbox \cite{VincentE2006ieeeaslp}. For this experiment, we run the models on four-source mixtures, and set $\lambda:\psi = 1:4$. \figref{fig:bss} summarizes the results. 

Without any remix loss, the baseline fails in recovering certain instruments, i.e., \texttt{piano} and \texttt{guitars} in the Slakh experiments, and \texttt{vocals} and \texttt{bass} in the MUSDB18 case. We observe that the performance gap mostly comes from the SAR scores, while the SIR improvement is less drastic. This signifies the neural remixers' tendency to suppress artifacts as much as possible as the source-specific artifacts do not cancel out. Meanwhile, it also shows that the proposed remixing loss benefits general separation performance according to the SIR improvement. 

\section{Conclusion}
We introduced a neural remixing model that works directly on the music mixture instead of assuming separated source tracks. Instead of a conventional separator-remixer workflow, we integrated the two processes into an end-to-end neural remixer via joint optimization. 
Results on Slakh and MUSDB18 showed that our proposed joint learning of remixing and separation greatly reduces the artifact produced in the process of source separation. Therefore, our models achieved significant improvement in the remix quality. From the perspective of user interaction, we demonstrated that the relationship between the estimated remix and the intended one is reasonably correlated as opposed to that induced from the baseline model. Sound examples and source codes are available at \url{https://saige.sice.indiana.edu/research-projects/neural-remixer}


\bibliographystyle{IEEEtran}
\bibliography{mjkim}

\begin{thebibliography}{10}
\providecommand{\url}[1]{#1}
\csname url@samestyle\endcsname
\providecommand{\newblock}{\relax}
\providecommand{\bibinfo}[2]{#2}
\providecommand{\BIBentrySTDinterwordspacing}{\spaceskip=0pt\relax}
\providecommand{\BIBentryALTinterwordstretchfactor}{4}
\providecommand{\BIBentryALTinterwordspacing}{\spaceskip=\fontdimen2\font plus
\BIBentryALTinterwordstretchfactor\fontdimen3\font minus
  \fontdimen4\font\relax}
\providecommand{\BIBforeignlanguage}[2]{{%
\expandafter\ifx\csname l@#1\endcsname\relax
\typeout{** WARNING: IEEEtran.bst: No hyphenation pattern has been}%
\typeout{** loaded for the language `#1'. Using the pattern for}%
\typeout{** the default language instead.}%
\else
\language=\csname l@#1\endcsname
\fi
#2}}
\providecommand{\BIBdecl}{\relax}
\BIBdecl

\bibitem{YoshiiK2005inter}
K.~Yoshii, M.~Goto, and H.~Okuno, ``{INTER}:{D}: a drum sound equalizer for
  controlling volume and timbre of drums,'' in \emph{The 2nd European Workshop
  on the Integration of Knowledge, Semantics and Digital Media Technology
  (EWIMT)}, 2005.

\bibitem{GilletO2005extraction}
O.~Gillet and G.~Richard, ``Extraction and remixing of drum tracks from
  polyphonic music signals,'' in \emph{Proc. of the IEEE International
  Conference on Acoustics, Speech, and Signal Processing (ICASSP)}, 2005.

\bibitem{WoodruffJ2006remixing}
J.~F. Woodruff, B.~Pardo, and R.~B. Dannenberg, ``Remixing stereo music with
  score-informed source separation.'' in \emph{Proc. of the International
  Society for Music Information Retrieval Conference (ISMIR)}, 2006.

\bibitem{ItoyamaK2008instrument}
K.~Itoyama, M.~Goto, K.~Komatani, T.~Ogata, and H.~G. Okuno, ``Instrument
  equalizer for query-by-example retrieval: Improving sound source separation
  based on integrated harmonic and inharmonic models.'' in \emph{Proc. of the
  International Society for Music Information Retrieval Conference (ISMIR)},
  2008.

\bibitem{Manilow2020hierarchical}
E.~Manilow, G.~Wichern, and J.~{Le Roux}, ``Hierarchical musical instrument
  separation,'' in \emph{Proc. of the International Society for Music
  Information Retrieval Conference (ISMIR)}, 2020.

\bibitem{Brocal2020content}
G.~{Meseguer-Brocal} and G.~Peeters, ``Content based singing voice source
  separation via strong conditioning using aligned phonemes,'' in \emph{Proc.
  of the International Society for Music Information Retrieval Conference
  (ISMIR)}, 2020.

\bibitem{StoterF2019open}
F.-R. St{\"o}ter, S.~Uhlich, A.~Liutkus, and Y.~Mitsufuji, ``Open-unmix-a
  reference implementation for music source separation,'' \emph{Journal of Open
  Source Software}, vol.~4, no.~41, p. 1667, 2019.

\bibitem{HuangP2015ieeeacmaslp}
P.-S. Huang, M.~Kim, M.~Hasegawa-Johnson, and P.~Smaragdis, ``Joint
  optimization of masks and deep recurrent neural networks for monaural source
  separation,'' \emph{IEEE/ACM Transactions on Audio, Speech, and Language
  Processing}, vol.~23, no.~12, pp. 2136--2147, Dec 2015.

\bibitem{UhlichS2017improving}
S.~Uhlich \emph{et~al.}, ``Improving music source separation based on deep
  neural networks through data augmentation and network blending,'' in
  \emph{Proc. of the IEEE International Conference on Acoustics, Speech, and
  Signal Processing (ICASSP)}, 2017.

\bibitem{SeetharamanP2019class}
P.~Seetharaman, G.~Wichern, S.~Venkataramani, and J.~{Le Roux},
  ``Class-conditional embeddings for music source separation,'' in \emph{Proc.
  of the IEEE International Conference on Acoustics, Speech, and Signal
  Processing (ICASSP)}, 2019.

\bibitem{LeeJ2019audio}
J.~H. Lee, H.-S. Choi, and K.~Lee, ``Audio query-based music source
  separation,'' \emph{Proc. of the International Society for Music Information
  Retrieval Conference (ISMIR)}, 2019.

\bibitem{SeetharamanP2019bootstrapping}
P.~Seetharaman, G.~Wichern, J.~{Le Roux}, and B.~Pardo, ``Bootstrapping deep
  music separation from primitive auditory grouping principles,''
  \emph{Workshop on Self-supervision in Audio and Speech at International
  Conference on Machine Learning}, 2019.

\bibitem{HersheyJ2016icassp}
J.~R. Hershey, Z.~Chen, J.~{Le Roux}, and S.~Watanabe, ``Deep clustering:
  Discriminative embeddings for segmentation and separation,'' in \emph{Proc.
  of the IEEE International Conference on Acoustics, Speech, and Signal
  Processing (ICASSP)}, 2016.

\bibitem{StollerD2018waveunet}
D.~Stoller, S.~Ewert, and S.~Dixon, ``{Wave-U-Net}: A multi-scale neural
  network for end-to-end audio source separation,'' in \emph{Proc. of the
  International Society for Music Information Retrieval Conference (ISMIR)},
  2018.

\bibitem{Lluis2019}
F.~Llu\'is, J.~Pons, and X.~Serra, ``{End-to-End Music Source Separation: Is it
  Possible in the Waveform Domain?}'' in \emph{Proc. of the Annual Conference
  of the International Speech Communication Association (Interspeech)}, 2019.

\bibitem{DefossezA2021mss}
D{\'e}fossez \emph{et~al.}, ``{Music Source Separation in the Waveform
  Domain},'' \emph{arXiv preprint arXiv:1911.13254}, 2019.

\bibitem{LuoY2019conv-tasnet}
Y.~Luo and N.~Mesgarani, ``{Conv-TasNet}: Surpassing ideal time--frequency
  magnitude masking for speech separation,'' \emph{IEEE/ACM Transactions on
  Audio, Speech, and Language Processing}, vol.~27, no.~8, pp. 1256--1266,
  2019.

\bibitem{YangH2021sanac}
H.~Yang, K.~Zhen, S.~Beack, and M.~Kim, ``Source-aware neural speech coding for
  noisy speech compression,'' in \emph{Proc. of the IEEE International
  Conference on Acoustics, Speech, and Signal Processing (ICASSP)}, 2021.

\bibitem{BryanN2013icml}
N.~J. Bryan and G.~J. Mysore, ``An efficient posterior regularized latent
  variable model for interactive sound source separation,'' in \emph{Proc. of
  the International Conference on Machine Learning (ICML)}, 2013.

\bibitem{ManilowE2019slakh}
E.~Manilow, G.~Wichern, P.~Seetharaman, and J.~{Le Roux}, ``Cutting music
  source separation some {Slakh}: A dataset to study the impact of training
  data quality and quantity,'' in \emph{Proc. of the IEEE Workshop on
  Applications of Signal Processing to Audio and Acoustics (WASPAA)}, 2019.

\bibitem{musdb18-hq}
\BIBentryALTinterwordspacing
Z.~Rafii, A.~Liutkus, F.-R. St\"oter, S.~I. Mimilakis, and R.~Bittner,
  ``{MUSDB18-HQ - an uncompressed version of MUSDB18},'' Aug. 2019. [Online].
  Available: \url{https://doi.org/10.5281/zenodo.3338373}
\BIBentrySTDinterwordspacing

\bibitem{VincentE2006ieeeaslp}
E.~Vincent, C.~Fevotte, and R.~Gribonval, ``Performance measurement in blind
  audio source separation,'' \emph{IEEE Transactions on Audio, Speech, and
  Language Processing}, vol.~14, no.~4, pp. 1462--1469, 2006.

\bibitem{ChoiHS2018phase-aware}
H.-S. Choi \emph{et~al.}, ``Phase-aware speech enhancement with deep complex
  {U-Net},'' in \emph{Proc. of the International Conference on Learning
  Representations (ICLR)}, 2018.

\bibitem{SawataR2020cumx}
R.~Sawata, S.~Uhlich, S.~Takahashi, and Y.~Mitsufuji, ``All for one and one for
  all: Improving music separation by bridging networks,'' in \emph{Proc. of the
  International Society for Music Information Retrieval Conference (ISMIR)},
  2020.

\bibitem{ParienteM2020asteroid}
M.~Pariente \emph{et~al.}, ``{Asteroid: The PyTorch-Based Audio Source
  Separation Toolkit for Researchers},'' in \emph{Proc. of the Annual
  Conference of the International Speech Communication Association
  (Interspeech)}, 2020.

\bibitem{YuD2017pit}
D.~Yu, M.~Kolb{\ae}k, Z.-H. Tan, and J.~Jensen, ``Permutation invariant
  training of deep models for speaker-independent multi-talker speech
  separation,'' in \emph{Proc. of the IEEE International Conference on
  Acoustics, Speech, and Signal Processing (ICASSP)}, 2017.

\bibitem{KingmaD2015adam}
D.~Kingma and J.~Ba, ``Adam: A method for stochastic optimization,'' in
  \emph{Proc. of the International Conference on Learning Representations
  (ICLR)}, 2015.

\bibitem{LeRouxJL2018sisdr}
J.~{Le Roux}, S.~Wisdom, H.~Erdogan, and J.~R. Hershey, ``{SDR --} half-baked
  or well done?'' in \emph{Proc. of the IEEE International Conference on
  Acoustics, Speech, and Signal Processing (ICASSP)}, 2019.

\bibitem{WierstorfH2017perceptual}
H.~Wierstorf \emph{et~al.}, ``Perceptual evaluation of source separation for
  remixing music,'' in \emph{Audio Engineering Society Convention 143}, 2017.

\end{thebibliography}

%
%
%
%
%

\end{document}